\documentstyle[twocolumn,aps,psfig]{revtex}
\begin{document}
\draft
\twocolumn[\hsize\textwidth\columnwidth\hsize\csname @twocolumnfalse\endcsname
%
%

\title{ One particle spectral weight of the three dimensional single band
Hubbard model}

\author{M. Ulmke$^{1*}$,
R. T. Scalettar$^1$, A. Nazarenko$^2$, and E. Dagotto$^2$}
            
\address{$1.$ Department of Physics, University of California, Davis, CA 95616}
\address{$2.$ Department of Physics, and National High Magnetic Field Lab,
Florida State University, Tallahassee, FL 32306}

\date{\today}
\maketitle

\begin{abstract} 
Dynamic properties of the three-dimensional single-band Hubbard model
are studied using Quantum Monte Carlo combined with the maximum
entropy technique.  At half-filling, there
is a clear gap in the density of states and well-defined
quasiparticle peaks at the top (bottom) of the lower (upper) Hubbard band.  
We find an antiferromagnetically induced weight above the 
naive Fermi momentum.
Upon hole doping, the chemical potential $\mu$  moves to
the top of the lower band where a robust peak is observed.
Results are compared with spin-density-wave (SDW) mean-field 
and self consistent Born approximation results, and also with
the infinite dimensional $(D=\infty)$ Hubbard model,
and experimental photoemission (PES) for three dimensional
transition-metal oxides. 

\end{abstract}

\vskip2pc]
\narrowtext

%
%


\section{Introduction}

The single band two dimensional 
Hubbard Hamilto\-nian\cite{HUBBARD} 
has recently received considerable attention due to 
possible connections with high temperature superconductors.
Indeed, evidence is accumulating that this Hamiltonian may describe, at least
qualitatively, some of the normal state properties 
of the cuprates.\cite{review}  
Exact Diagonalization (ED) and Quantum Monte Carlo (QMC) have been
used to model static properties
like the behavior of spin correlations and magnetic
susceptibility both at half-filling
and with doping.\cite{review} 
Comparisons of dynamic quantities like the spectral weight
and density of states with
angle-resolved photoemission results\cite{flat-exper,flat,bulut,hanke,berlin}
have also proven quite successful.
Significantly, while analytic calculations have pointed towards 
various low temperature 
superconducting instabilities, such indications have been 
absent in numerical work.\cite{review}

Historically, however, the Hubbard model was first proposed
to model magnetism and metal-insulator transitions 
in 3D transition metals and their oxides,\cite{HUBBARD}
rather than superconductivity.  Now that the technology of numerical
work has developed, it is useful to reconsider some of these
original problems.  A discussion of possible links between  
the 3D Hubbard model and photoemission results for
${\rm YTiO_3}$, ${\rm Sr VO_3}$ and others\cite{fujimori,inoue,morikawa}
has already recently occurred.  In such 
perovskite ${\rm Ti^{3+}}$ and ${\rm V^{4+}}$ oxides, which are both in a 
$3d^1$ configuration, 
the hopping amplitude $t$ between transition-metal
ions can be varied by
modifying the $d-d$ neighboring overlaps through a tetragonal distortion. 
Thus, the strength of the
electron correlation $U/t$ can be varied by changing the composition.
In fact, a metal-insulator
transition has been reported in the series 
${\rm SrVO_3}-{\rm Ca VO_3}-{\rm La Ti O_3}-{\rm YTiO_3}$.
On the metallic side, a quasiparticle band is experimentally 
observed near the
Fermi energy $E_F$, as well as a high energy satellite associated to the
lower Hubbard band (LHB).\cite{fujimori,rrmp}
Spectral weight is transferred from the quasiparticle to the LHB as
$U/t$ is increased at half-filling.

In this paper, we report the first use of Quantum Monte Carlo, combined with
analytic continuation techniques, to evaluate the spectral
function and density of states for the 3D Hubbard
Hamiltonian.  The motivation is twofold.  First, we want
to compare general properties of the 3D Hubbard Hamiltonian
with the extensive studies already reported in 
the literature\cite{WHITE,jarrell,rrmp,review} for the 2D and infinite-D cases.
Of particular importance is the presence of quasiparticles
near the half-filling regime, as well as the evolution of spectral
weight with doping.  Many of the high-Tc cuprates contain ${\rm CuO_2}$
planes that are at least weakly coupled with each other, and thus the
study of the 3D system may help in understanding part of the details of
the cuprates.  More generally,
the Hubbard Hamiltonian is likely to continue being one 
of the models used to capture the physics of 
strongly correlated electrons, so
we believe it is important to document its
properties in as many environments as possible for potential 
future comparisons against experiments. 

Secondly, we discuss a particular illustration
of such contact between Hubbard Hamiltonian
physics and experiments on 3D transition metal oxides.
In addition to the studies of half-filled systems with varying
correlation energy mentioned above,
experiments where the band filling is tuned by changing the chemical
composition have also been reported.\cite{fujimori2,morikawa,tokura} 
One compound that has been carefully
investigated in this context is ${\rm Y_{1-x}
Ca_x Ti O_3}$. At $x=0$ the system is an
antiferromagnetic insulator. As $x$ increases, a
metal-insulator
transition is observed in PES studies. The lower and upper Hubbard bands
(LHB and UHB) are easily identified
even with $x$ close to 1, which would naively correspond to small
electronic density in the single band Hubbard model, i.e. a regime where
$U/t$ is mostly irrelevant. In the experiments,
a very small amount of spectral weight is
transferred to the Fermi energy, filling the gap observed at half-filling
(i.e.  generating a ``pseudogap'').

Analysis of the PES
results of these compounds using the paramagnetic solution of
the Hubbard Hamiltonian in
infinite-D \cite{metzner}, a limit where dynamic mean field theory becomes
exact (see section II), has resulted in qualitative agreement
\cite{jarrell,georges,rrmp} with the experimental results.
At and close to half-filling there is an antiferromagnetic (AF)
solution which becomes unstable against a paramagnetic (PM) solution at
a critical concentration of holes. In the PM case, weight
appears in the original Hubbard gap as reported experimentally.
However, this analysis of the spectral
weight in terms of the infinite-D Hamiltonian
is in contradiction with results for the density of
states reported in the 2D
Hubbard model\cite{review} where it is found that upon hole (electron) 
doping away from half-filling the chemical potential $\mu$ moves to the top 
(bottom) of the valence (conduction) band. The results at $\langle n
\rangle =1$ in 2D already show the presence of a robust quasiparticle
peak which is absent in the insulating PM solution of the $D=\infty$ model.  
That is, in the 2D system the large peak in the density of states observed away
from half-filling seems to evolve from a robust peak already present at
half-filling. On the other hand, at $D=\infty$ a feature resembling a
``Kondo-resonance'' is $generated$ upon doping if
the paramagnetic solution is used.  This peak in the density
of states does not have an analog at half-filling unless
frustration is included.\cite{jarrell} 
Studies in 3D may help in the resolution of this apparent 
non-continuity of the physics of the Hubbard model when the dimension 
changes from 2 to $\infty$.
The proper way to carry out
a comparison between $D=3$ and $\infty$ features is to base
the analysis on ground state properties. With this
restriction, i.e. using the AF solution at $D=\infty$ and
close to half-filling, rather than the PM solution,
we found that the $D=3$ and $\infty$ results are in good
agreement.

In this paper we will consider which of these situations the
3D Hubbard Hamiltonian better corresponds to, and therefore 
whether the single band Hubbard Hamiltonian provides an
adequate description of the density of states of 3D transition-metal oxides.

\section{Model and Methods}
The single band Hubbard Hamiltonian is
\begin{eqnarray}
H & = & -t \sum_{\bf \langle ij \rangle } ( c^\dagger_{ {\bf i} \sigma}
c_{{\bf j} \sigma} + h.c.)  
- \mu \sum_{{\bf i}\sigma} n_{{\bf i}\sigma} \nonumber \\
& & + U \sum_{\bf i} (n_{{\bf i} \uparrow} - 1/2 )
(n_{{\bf i} \downarrow} - 1/2 ),
\label{hubbard}
\end{eqnarray}
where the notation is standard. Here ${\bf \langle ij \rangle }$
represents nearest-neighbor links on a 3D cubic lattice. 
The chemical potential $\mu$ controls the doping. For $\mu=0$ the
system is at half filling ($\langle n \rangle=1$) due to particle-hole 
symmetry. $t\equiv 1$ will set our energy scale. 

We will study the 3D Hubbard Hamiltonian using a finite temperature,
grand canonical Quantum Monte Carlo (QMC) method \cite{blankenbecler} which
is stabilized at low temperatures by the use of orthogonalization techniques
\cite{white}. The algorithm is based on a functional-integral representation
of the partition function obtained
by discretizing the ``imaginary-time'' interval
$[0,\beta]$ where $\beta$ is the inverse temperature. The Hubbard interaction
is decoupled by a two-valued Hubbard-Stratonovich transformation \cite{hirsch}
yielding a bilinear time-dependent fermionic action. The fermionic degrees of
freedom can hence be integrated out analytically, and the partition function
(as well as observables) can be written as a sum over the auxiliary fields 
with a weight proportional to the product of two determinants, one for 
each spin species. 
At half-filling ($\langle n \rangle=1$),
it can be shown by  particle-hole transformation of one spin species
$(c_{{\bf i}\downarrow} \rightarrow 
(-1)^{{\bf i}} c_{{\bf i}\downarrow}^\dagger)$ 
that the two determinants differ only by a positive factor, hence their 
product is positive definite. 
At general fillings, however, the product can become negative,
and this ``minus-sign problem'' restricts the application of QMC
to relatively high temperature (of order 1/30 of the bandwidth)
off half-filling.  

The QMC algorithm provides a variety of static and dynamic observables. 
One equal time quantity in which we are interested is
the magnetic (spin-spin) correlation function,
\begin{equation}
C({\bf l}) = \frac{1}{N} \sum_{{\bf j}} \langle m_{\bf j} m_{{\bf j+l}} \rangle.
\label{correl} 
\end{equation}
Here $m_{\bf j}=\sum_\sigma\sigma n_{{\bf j}\sigma}$ is the local 
spin operator, and $N$ is the total number of lattice sites.
Static correlations have also been investigated in earlier
studies of the $3D$ Hubbard model \cite{HIRSCHn,rts} where the 
antiferromagnetic phase diagram at half filling was explored. 

To obtain dynamical quantities in real time or frequency, the 
QMC results in imaginary time have to be analytically continued 
to the real time axis. Since we are mostly interested in the one-particle
spectrum we measure the one-particle Green function $G({\bf p},\tau)$.
The imaginary part of $G({\bf p},\omega)$ (in real frequency) defines
the spectral weight function at momentum ${\bf p}$, 
$A({\bf p},\omega)$, which
is related to $G({\bf p},\tau)$ by:
\begin{equation}
G({\bf p},\tau) = \int_{-\infty}^\infty d\omega \, A({\bf p},\omega) \, 
\frac{e^{-\tau\omega}}{1+e^{\beta\omega}}.
\label{conti}
\end{equation}
$A({\bf p},\omega)$ can in principle be calculated by inverting 
Eq.(\ref{conti}), but
the exponential behavior of the kernel at large values of $|\omega|$
makes this inversion difficult numerically.
$G({\bf p},\tau)$ is quite insensitive 
to details of $A({\bf p},\omega)$ in particular at large frequencies. 
Since $G({\bf p},\tau)$ is known only on a finite grid in the interval 
$[0,\beta]$
and there only within the statistical errors given by the QMC-sampling,
solving Eq.(\ref{conti}) for $A({\bf p},\omega)$ is an  ill-posed problem.
A large number of solutions exists, and the problem is to find criteria
to select out the correct one. 
This can be done by employing the Maximum Entropy (ME) method 
\cite{gubernatis}. Basically, the ME finds the ``most likely'' solution
$A({\bf p},\omega)$ which is consistent with the data and all information 
that is known about the solution (like positivity, normalization, etc.).
ME avoids ``overfitting'' to the data by a ``smoothing'' technique 
that tries to assimilate the resulting $A({\bf p},\omega)$ to a flat default
model. In the absence of any data ($G({\bf p},\tau)$) ME would converge to the 
default model which is chosen to be a constant within some large frequency
interval. There is no adjustable parameter in the ME application. 
One needs accurate data for $G({\bf p},\tau)$ with a statistical 
error of $o(10^{-4})$ to get reliable results for $A({\bf p},\omega)$.   
  
In principle, one can calculate analytically the first and second (and higher)
moments of the spectral weight and include this information in the
ME procedure. However, we chose to calculate the moments afterwards from
the resulting  function $A({\bf p},\omega)$, and to compare them with the 
analytically known results as a further test. The agreement was in all 
cases within 10\%.
Still, the ME methods provides only a rough estimate of the true spectral
weight functions. Band gaps and the positions of significant peaks are 
usually well captured but fine structure which needs a high frequency
resolution is hard to detect within the ME approach.

Integrating $A({\bf p},\omega)$ over the momenta gives the one-particle 
density of states (DOS) $N(\omega)$. However, technically, 
it is preferrable to 
integrate first $G({\bf p},\tau)$, which reduces the statistical errors, and 
then perform the analytic continuation. 

The DOS will be compared to results from the dynamical
mean-field theory of infinite dimensions, $D=\infty,$ \cite{metzner}. 
In this limit, with the proper scaling of the hopping element 
($t=t^*/\sqrt{Z}$, with $Z$ being the coordination number) the one-particle 
self energy becomes local or, equivalently, momentum independent and the 
lattice problem is mapped onto a single site problem.
The constant $t^*$ is set to $t^*=\sqrt{6}$ to obtain the same energy scale,
$t=1$, when compared to the $3D$ case. \cite{bethe} 
In contrast to conventional mean-field theories, the self energy remains 
frequency dependent, preserving important physics. 
Spatial fluctuations are neglected, an approximation
which becomes exact in the limit
$Z\to\infty$ ($Z=2D$ for the simple cubic lattice).
Even in $D=\infty$, the remaining local interacting problem cannot be solved
analytically but will also be treated by a finite temperature QMC
\cite{fye} supplemented by a self-consistency iteration 
\cite{georges,rrmp}. The advantage is that the system can be investigated in the
thermodynamic limit with a modest amount of computer time. 
Due to its local character the $D=\infty$ approach cannot provide information
on momentum dependent spectral functions.
However, recently a $k$-resolved
spectral function has been studied\cite{rspec} in $D=\infty$. 

Among other things, the $D=\infty$ limit has been used recently to study 
the AF phase diagram in the Hubbard model \cite{jarrell}. 
The agreement of the N\'eel temperature with 3D
results is good.\cite{rts} In $D=\infty$ it is further possible 
to suppress AF long range order artificially by restricting 
the calculation to the 
(at low temperatures unstable) paramagnetic solution at half-filling. 
In this way, one may simulate frustration due to the lattice structure 
or orbital degeneracy, although in the absence of
calculations for hypercubic lattices with nearest and
next-nearest uniform hopping amplitudes
it is still a conjecture how close this approach is to including these
effects fully.
 
\section{Half-filling}

\subsection{Quantum Monte Carlo}

We first study the single particle spectral weight $A({\bf p},\omega)$ 
at relatively strong coupling, $U=8$, and half-filling ($\langle n \rangle=1$)
at a low temperature of $T=1/10$.
%
\begin{figure}
\psfig{file=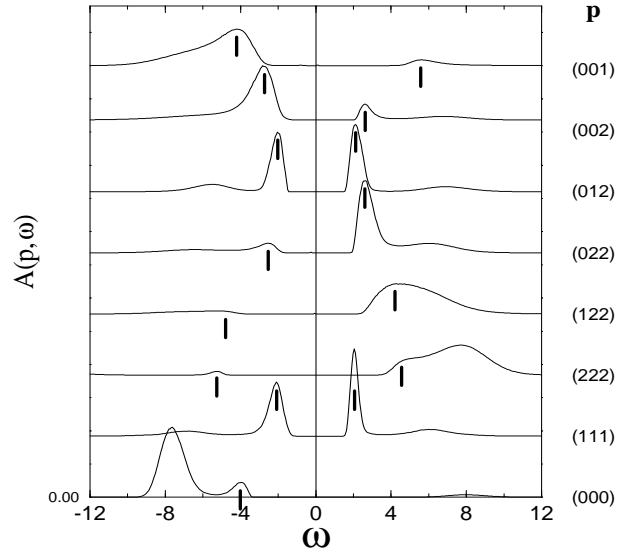,height=3.5in,width=4.in}
\vskip-10mm
\caption{Single particle spectral function $A({\bf p},\omega)$ versus
$\omega$ for several momenta. The results correspond to a lattice
with $4^3$ sites, $U=8$, $\beta=10$, half-filling and using $t\equiv1$ 
as energy scale. Momenta, $\bf p$, are in units of $\pi/2$. Bars indicate
the position of the quasiparticle peak.}

\end{figure}
A gap is clearly present in the spectrum (Fig.~1) which is
compatible with the expectation that the half-filled Hubbard model on a 
bipartite lattice is
an antiferromagnetic insulator for all nonzero values of the coupling 
$U/t$. The
spectral weight has  four distinct features (two in the
LHB and two identical ones in the UHB, as expected from particle-hole 
symmetry).
In the UHB there is weight at a high energy, roughly in the interval between 
$\sim
5t$ and $8t$. This broad feature likely corresponds to the incoherent part of
the spectral function found in previous simulations for the 2D Hubbard
and $t-J$ models.\cite{review}
The dominant scale of this incoherent weight is $t$, 
and since it is located far
from the top of the valence band its presence is not important for the
low temperature properties of the system.

Much more interesting is the sharper peak found close to the 
gap in the spectrum. This band dispersion starts at a binding energy of
approximately $\omega= -4t$ at
momenta $(0,0,0)$ and moves up in energy
obtaining its maximum value at $\omega\approx -2t$
at momenta $(0,\pi/2,\pi)$ and
$(\pi/2,\pi/2,\pi/2)$ in Fig.~1. The width of the peak diminishes as the
top of the valence band is reached. Similar structure was discussed
before in studies of 2D systems, which had a somewhat higher resolution,
as a ``quasiparticle'' band corresponding to a hole moving coherently
in an antiferromagnetic background.\cite{review,moreo} This quasiparticle
should be visualized as a hole distorting the AF order 
parameter in its vicinity. In this respect it is like a spin-polaron or
spin-bag,\cite{schrieffer} although ``string states'' likely influence
its dispersion and shape.\cite{review}
The quasiparticle (hole plus spin distortion) movement is
regulated by the exchange $J$, rather than $t$.

Using the center of the quasiparticle peaks of Fig.~1
as an indication of the actual quasiparticle pole position, we obtain a 
 bandwidth $W$ of about $2$ to $3t$ or,
equivalently, $4$ to $6J$ using ${ J=4t^2/U}$ for $U=8$.
However, due to the low resolution of the ME
procedure, reflected in part in the large width of the peaks of Fig.~1,
it is difficult to show more convincingly within QMC/ME that the
quasiparticle bandwidth is indeed dominated by $J$. 

Note that moving from $(0,0,0)$ to $(\pi,\pi,\pi)$ along the main
diagonal of the Brillouin Zone (BZ), the PES part of the spectrum 
(i.e. the weight at $\omega <0$) loses intensity.
There is a clear
transfer of weight from PES at small $|{\bf p}|$ to IPES at large $|{\bf
p}|$, as observed in 2D simulations.\cite{bulut1} In addition,
note that there is PES weight above the (naive) Fermi momentum of this
half-filled system. For example, at ${\bf p} = (0,\pi,\pi)$, spectral
weight at $\omega <0$ can be clearly observed. Similarly, at 
${\bf p} = (0,0,\pi)$
weight in the IPES region is found.  This effective 
doubling of the size of the unit cell in all three directions
is a consequence of the presence of AF long range order.
The hole energy at ${\bf p}=(0,0,0)$ and
$(\pi,\pi,\pi)$ becomes degenerate in the bulk limit and
the quasiparticle band, for example along the main diagonal 
of the BZ,
has a reflection symmetry with respect to $(\pi/2,\pi/2,\pi/2)$, as
observed in our results (Fig.~1). 
However, note
that the actual intensity of the AF-induced PES weight close to
$(\pi,\pi,\pi)$ is a function of the coupling. As $U/t \rightarrow 0$, the
intensity of the AF-induced region is also reduced to zero. 
The presence of this AF-generated feature  
has recently received
attention in the context of the 2D high-Tc cuprates.\cite{schrieffer,shadow} 
While its presence
in PES experiments at optimal doping is still under discussion, these features
clearly appear in PES experimental studies of half-filled
insulators, like ${\rm Sr_2 Cu O_2 Cl_2}$.\cite{wells}
Thus, while this behavior has been primarily discussed in the context
of 2D systems angle-resolved PES (ARPES) studies of 3D insulators
like ${\rm LaTiO_3}$ might also show such features.

In Fig.~2 we show the density of states (DOS) $N(\omega)$ of the $4^3$ lattice.
The two features described before, namely quasiparticle and incoherent 
background, in both PES and IPES are clearly visible.
Also shown in Fig.~2 is the temperature effect on $N(\omega)$
which is weak for the given temperatures ($\beta=10$ and 4).
The basic features are still retained,
only the gap is slightly reduced and the quasiparticle peak less pronounced
at the higher temperature.

Results corresponding to a larger coupling, $U=12$, at $\beta=4$ are shown 
in Fig.~3.
The gap increases and the quasiparticle band becomes sharper as $U/t$ grows, 
as expected if its bandwidth is regulated by $J$. $A({\bf p},\omega)$ at
$U=12$, $\beta=4$ are similar to the results shown in Fig.~1.

\begin{figure}
\psfig{file=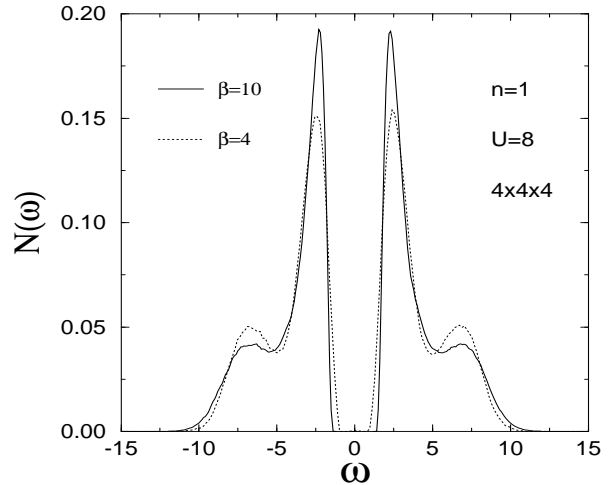,height=3.5in,width=4.2in}
\vskip-20mm
\caption{
Density of states $N(\omega)$ of the 3D Hubbard model on a $4^3$
lattice, $U=8$, $\beta=10$ (solid line) and $\beta=4$ (dashed line) and 
$\langle n \rangle=1$.  We do not enforce the $\rho(\omega)=\rho(-\omega)$
symmetry which occurs at half-filling due to particle-hole symmetry.
However, deviations from this constraint are small, a further check on
our numerics.}

\end{figure}
\begin{figure}
\psfig{file=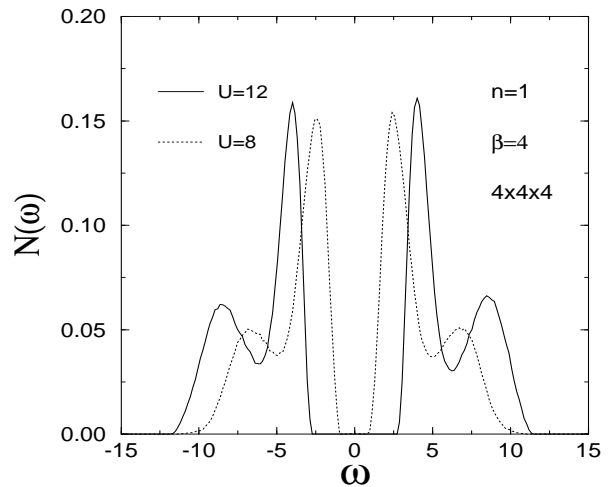,height=3.5in,width=4.2in}
\vskip-20mm
\caption{Density of states $N(\omega)$ of the 3D Hubbard model on a $4^3$
lattice at $U=8$ (dashed line) and $U=12$ (solid line), $\beta=4$, and 
$\langle n \rangle=1$.}

\end{figure}

A characteristic double peak like that seen
in Figs.~2,3 has been observed in the X-ray absorption
spectrum of ${\rm LaFeO_3}$ \cite{sarma}, which is a strongly correlated,
wide gapped antiferromagnetic insulator. The peaks appear at energies of  
about 2.2 eV and 3.8 eV. 
When Fe is substituted by Ni this structure vanishes and the system 
becomes a paramagnetic metal at a Ni concentration of about 80\%.
If we choose for comparison with the calculated DOS (Fig.~3) 
a hopping amplitude of $t=0.5$ eV, giving a reasonable d-bandwidth of $W=6$ eV,
the positions of the quasiparticle peak and the maximum of the incoherent
band for $U=12t$ are at about $\omega_1\approx 4t=2$ eV and 
$\omega_2\approx 8.4t=4.2$ eV, respectively. 
The agreement with the experimental values
is fairly good considering the crude simplifications of the Hubbard model 
such as neglecting orbital degeneracy and charge transfer effects. 
Even the estimated charge gap of Fig.~3, defined by the onset of  
spectral weight relative to the Fermi energy,   
$\Delta_{charge}\approx 3t=1.5$ eV is not too 
far from the experimental value of $\sim 1.1$ eV.
The ratio $r=\omega_2/\omega_1$ decreases with $U$ since for $U\gg t$
both energies are expected to converge to $U/2$.
While for $U=12t$, $r\approx 2.1$ is comparable to the
experimental value ($\sim 1.7$), it is too large for $U=8$ ($r \approx 2.9 $)
showing that under the assumption of  
a single band Hubbard model description for ${\rm LaFeO_3}$, 
the effective on-site interaction is at 
least of the size of the d-bandwidth.  

Another feature which has been attributed to antiferromagnetic ordering
was found in a high resolution PES study of $\rm{V_2O_3}$. \cite{shin,rozen2}
In the AF insulator at $T=100K$ the spectrum shows a shoulder at 
$\omega_1=-0.8$ eV which is absent in the paramagnetic metal at $T=200K$. 
This shoulder might be a reminiscent of the quasiparticle peak.
The maximum of the lower Hubbard band is at about $\omega_2 \approx -1.3$ eV, 
giving a ratio $\omega_2/\omega_1\approx 1.6$ similar to that observed in
 ${\rm LaFeO_3}$. The on-site interaction was estimated to be about 1.5 times 
the bandwidth.\cite{shin} 

\begin{figure}
\psfig{file=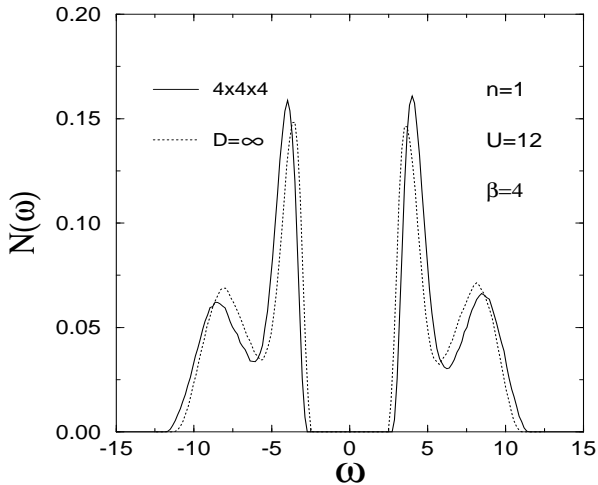,height=3.5in,width=4.2in}
\vskip-20mm
\caption{Density of states $N(\omega)$ of the 3D Hubbard model on a $4^3$
lattice at $U=12$,  $\beta=4$ and $\langle n \rangle=1$ (solid line)
compared with the density of states at the same parameters for 
$D=\infty$ (dashed line).}
\end{figure}

It is interesting to compare the results obtained in our simulations
with those found in the $D=\infty$ limit of the Hubbard model. At
half-filling for arbitrary coupling strength, the $D=\infty$
model has an AF insulating ground state. Its DOS is shown
in Fig.~4, using the same coupling and temperature as in the 3D
simulation.
$N(\omega)$ for both cases  are similar, and they are also similar
to results found before in 2D, suggesting that the physics of holes
in an antiferromagnetic system is qualitatively the same irrespective of
whether a 2D, 3D or
${\rm \infty}$D lattice is used, at least within the accuracy of
present QMC/ME simulations. 

\subsection{SDW mean-field and Born approximation}

Since the data shown in the previous subsection correspond to holes
in a system with AF long-range order, it is natural to
compare our results against those found in mean-field approximations to
the half-filled 3D Hubbard model that incorporate magnetic order
in the ground state. The ``spin-density-wave'' mean-field 
approximation has been extensively used in the context of the 2D Hubbard
model,\cite{sdw} and here we will apply it to our 3D problem. 
For a
lattice of $N$ sites, the self-consistent equation for the gap $\Delta$ is
\begin{equation}
1={{U}\over{2N}} \sum_{\bf p} {{1}\over{E_{\bf p}}},
\label{sdw}
\end{equation}
where $E_{\bf p} = \sqrt{ \epsilon_{\bf p}^2 + \Delta^2 }$ is the
quasiparticle energy, and $\epsilon_{\bf p} = -2t (\cos p_x + \cos p_y +
\cos p_z)$ is the bare electron dispersion. 
The resulting quasiparticle dispersion is shown in Fig.~5 compared
against the results of the QMC/ME simulation. The overall agreement is
good if the coupling $U$ in the gap equation Eq.(4) is tuned to a value
$U \sim 5.6$.  It is reasonable that a reduced $U$ should be required
for such a fit, since the SDW MF gap is usually larger than the more
accurate QMC result.  Similar renormalizations of $U$ in comparing
QMC and approximate analytic work has been discussed in the context
of fitting the magnetic response,\cite{renormu}
and has also been explicitly calculated.\cite{vandongen}
Fig.~5 shows many of the features observed in the numerical simulation,
namely a hole dispersion which is maximized at $(\pi/2,\pi/2,\pi/2)$ 
for the momenta shown there, an overall bandwidth
smaller than the noninteracting one, and the presence of AF-induced
features in the dispersion above the naive Fermi momentum.

Thus the SDW MF approach qualitatively captures the correct
hole quasiparticle bandwidth $J$ at half-filling.  
However, a spurious
degeneracy appears in the hole dispersion in this
approximation.  Momenta satisfying
$\cos p_x + \cos p_y + \cos p_z =0$ have the same energy. This is not
induced by symmetry arguments and is an artifact of the SDW approach.
In addition, $A({\bf p},\omega)$ in the SDW approximation only has
one peak in the PES region for each value of the momentum, missing
entirely the incoherent part.  

While it may not be
necessary to fix this problem in this case, it is important in
general to be able to go beyond SDW MF.  To do this,
the self-consistent Born approximation\cite{born} (SCBA)
for one hole in the 3D $t-J$ model, which corresponds to
the strong coupling limit of the
Hubbard model, can be used. This
technique reproduces accurately exact diagonalization
results in the 2D case.\cite{born} 

\begin{figure}
\vspace*{10mm}
\hskip3mm
\psfig{file=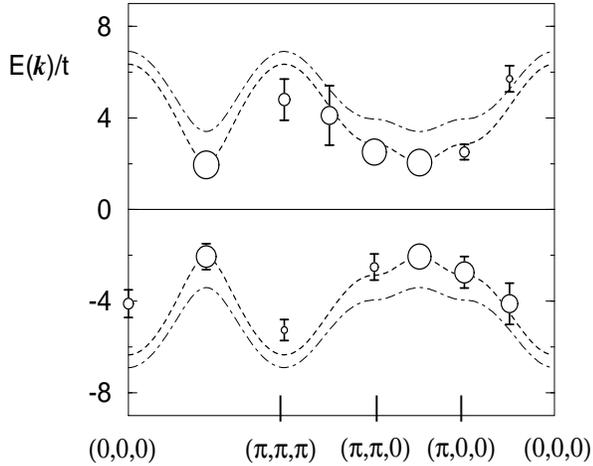,height=1.1in,width=1.9in,angle=-90}
\vskip28mm
\caption{Quasiparticle dispersion in the SDW MF approximation at
half-filling and zero temperature.
Results are at $U/t=8$ (dot-dashed line) and $U=5.57t$
(dashed line).
QMC/ME results for the 3D Hubbard
model on a $4^3$ lattice at the same coupling, density and temperature
are also shown (open circles). 
The area in the circles is proportional to the peak intensity.
The error bars are the width
of the peak. For some momenta the intensity of the PES or IPES data is
so low that no peak position is reported.}

\end{figure}
\begin{figure}
\vskip15mm
\hskip3mm
\psfig{file=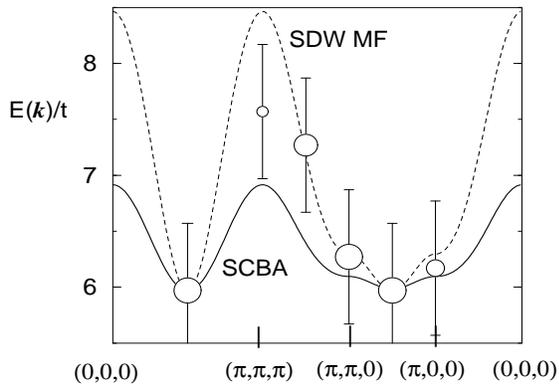,height=1.9in,width=1.9in,angle=-90}
\vskip-3mm
\caption{Quasiparticle dispersion in the SDW MF approximation at $U=12.8t$
(dashed line) compared against similar results obtained with the
SCBA approximation for one hole in the $t-J$ model at $J/t = 0.3125$
(solid line). The energy scale of the SCBA dispersion is shifted such that
the bottom of the band at $(\pi/2,\pi/2,\pi/2)$ agrees with the SDW MF
result.  The QMC data, also shown, lie between these weak and strong
coupling approximations.  
The area in the circles is proportional to the peak intensity.
The error bars are the width of the peak.}

\end{figure}

Actually, the dispersion of a dressed hole in an antiferromagnet 
within the SCBA
for a bilayer system, and also 
for a 3D cubic lattice, has been recently 
studied.\cite{sasha} Here, for completeness,
we reproduce some of the results of Ref.\cite{sasha}, and compare them 
against those of the 3D
Hubbard model obtained with the SDW MF approximation
and QMC calculations (Fig.~6). 
The comparison is carried out at $J/t \sim 0.3$
which corresponds to $U/t \sim 13$. The maximum 
of the dispersion in the valence band using the SCBA
now lies at $({\pi/2, \pi/2, \pi/2})$, removing the spurious SDW MF
degeneracy. In the scale of
Fig.~6 the splitting between this momentum and $(\pi,\pi/2,0)$ 
is difficult to resolve, since it  corresponds to about 100K.
Note that the bandwidth
predicted by the SDW MF technique is approximately a factor two larger
than the more accurate prediction of the SCBA.
However, for this larger value of $U$ it does not appear possible to fit 
simultaneously the
SDW MF bandwidth and band-gap to the results of QMC
by the same renormalization of $U$,
something which can be done successfully at weaker coupling, $U=8$. 
The QMC points at this intermediate coupling 
value where $U$=bandwidth lie in between the SDW MF and SCBA. 
Though the uncertainties in the QMC results are rather large,
we expect the agreement between SCBA and QMC results to improve
as the coupling increases.
The best fit of the SCBA data\cite{sasha} is
$\epsilon({\bf p})\,=\,c\,+\,0.082(\cos p_x\cos p_y+
\cos p_y\cos p_z+\cos p_x\cos p_z)\,
+\,0.022(\cos 2p_x+\cos 2p_y+\cos 2p_z)$ (eV), if $J=0.125eV$ and
$t=0.4eV$ are used. The constant $c$ is defined by the SDW MF gap (Fig.~6).
As in the case of the 2D problem,
holes tend to move within the same sublattice to avoid
distorting the AF background.\cite{review} Working 
at small $J/t$, the bandwidth of the
3D $t-J$ hole quasiparticle was found to scale 
as $J$,\cite{sasha} as occurs in two dimensions.

\section{Finite Hole Density}

\subsection{D=3}

We can also 
use the QMC approach to study the 3D Hubbard model away
from half-filling for temperatures down to about
1/30 of the bandwidth, a value for which 
$T\sim J$ for the present strong coupling values. 
First, we study the influence of doping and temperature on the 
spin-spin correlation function $C({\bf l})$. 
At half filling $C({\bf l})$ shows strong antiferromagnetic correlations over 
the whole $4^3$-lattice at $\beta=10$ (Fig.~7). 
At $\beta=2$ the correlations are significantly weakened, 
and with additional doping ($\langle n \rangle =0.88$)
all correlations are suppressed besides those between nearest neighbors.
These appear to be stable against doping.
The density of local moments, $\sqrt{C(0)}$ reaches its 
low temperature limit at 
an energy scale set by $U$ and hence is unaffected by the
change of $\beta$ from $\beta=2$ to $\beta=10$
(note that longer range spin correlations form at a temperature set by
the much smaller energy scale $J$).
$\sqrt{C(0)}$ is to first order proportional to the electronic
density and hence slightly reduced at $\langle n \rangle =0.88$.

There has been considerable discussion concerning the relationship
between the spin-spin correlations and the presence of a gap
in the density of states.  In particular, it was observed\cite{WHITE} that
if $N(\omega)$ is evaluated on lattices of increasing size at 
fixed temperature a well formed gap appearing on small
lattices disappears when the spatial extent exceeds the spin-spin
correlation length.  Decreasing the temperature (and hence increasing
the range of the spin correlation) allows the gap to reform.
Similar effects are seen here in 3D.  
\begin{figure}
\hspace*{0mm}
\psfig{file=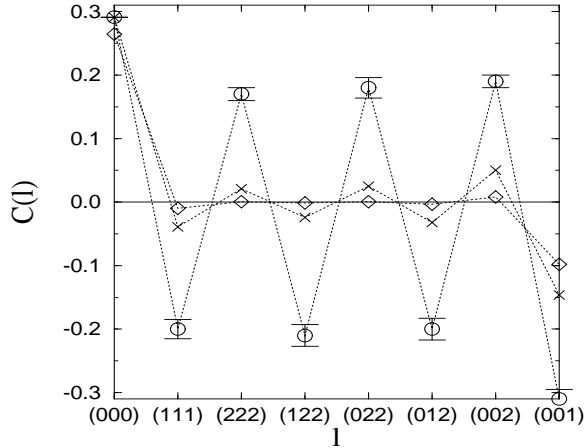,height=3.3in,width=4.0in}
\vskip-16mm
\caption{Spin-spin correlation function $C({\bf l})$ over a path in real space 
on the $4^3$ lattice at $U=8$, $\beta=10$, $\langle n \rangle=1$
($\circ$); $\beta=2$, $\langle n \rangle=1$ ($\times$);
and $\beta=2$, $\langle n \rangle=0.88$ ($\diamond$).
$C(0)$-values are divided by a factor of 3 for clearness.
Error bars are smaller than the symbols when not shown.}

\end{figure}

Fig.~8a shows the density of states on a $4^3$ lattice at several densities,
$U=8$ and $\beta=2$. 
At this temperature the charge gap is not fully developed, and the
quasiparticle peaks cannot be resolved.
The result with doping is similar to that reported on 2D
lattices.\cite{bulut2} The chemical potential $\mu$ moves to
the top of the valence band as the density is reduced from half-filling. 
A large peak is generated which increases in intensity as $\langle
n \rangle$ is further reduced. 
The weight of the upper part of the spectrum (reminiscent of the UHB)
decreases with doping due to the reduced effective interaction.
Similar results are shown in Fig.~8b but for a $6^3$ lattice. 
There is not much difference between the two lattices, showing that 
within the resolution of the ME procedure finite size effects are small.

The large peak that appears in Fig.~8a-b at
finite hole density is crossed by $\mu$ as the density is reduced. At
$\langle n \rangle = 0.94$, the peak is located to the left of $\mu$, at
$\langle n \rangle = 0.88$ it has reached the chemical potential, and at 
$\langle n
\rangle = 0.72$, the peak has moved to the right. This is in agreement
with the behavior observed in both 2D QMC and ED
simulations,\cite{dos} and it may be of relevance for estimations of
superconducting critical temperatures if a source of carrier attraction is
identified.\cite{dos1}

The results of the previous section
at half-filling obtained at low temperatures $(T\sim 1/10)$
revealed a sharp quasiparticle peak in the DOS at the top of the valence
band and bottom of the conduction band. 
Numerical studies of
2D lattices have shown that the peak intensity at $T=0$ 
is the $largest$ at half-filling.\cite{dos} Away from half-filling, the peak is
still visible but it is broader than at $\langle n \rangle =1$.\cite{dos} 
Thus there is no evidence that the sharp peak in the DOS of
the doped system has been generated dynamically and represents 
a ``Kondo-resonance'' induced by doping,
as has sometimes been suggested \cite{jarrell}, and as Fig.~8 obtained at
relatively high temperature, $\beta=2$, seem to imply.

\begin{figure}
%
\psfig{file=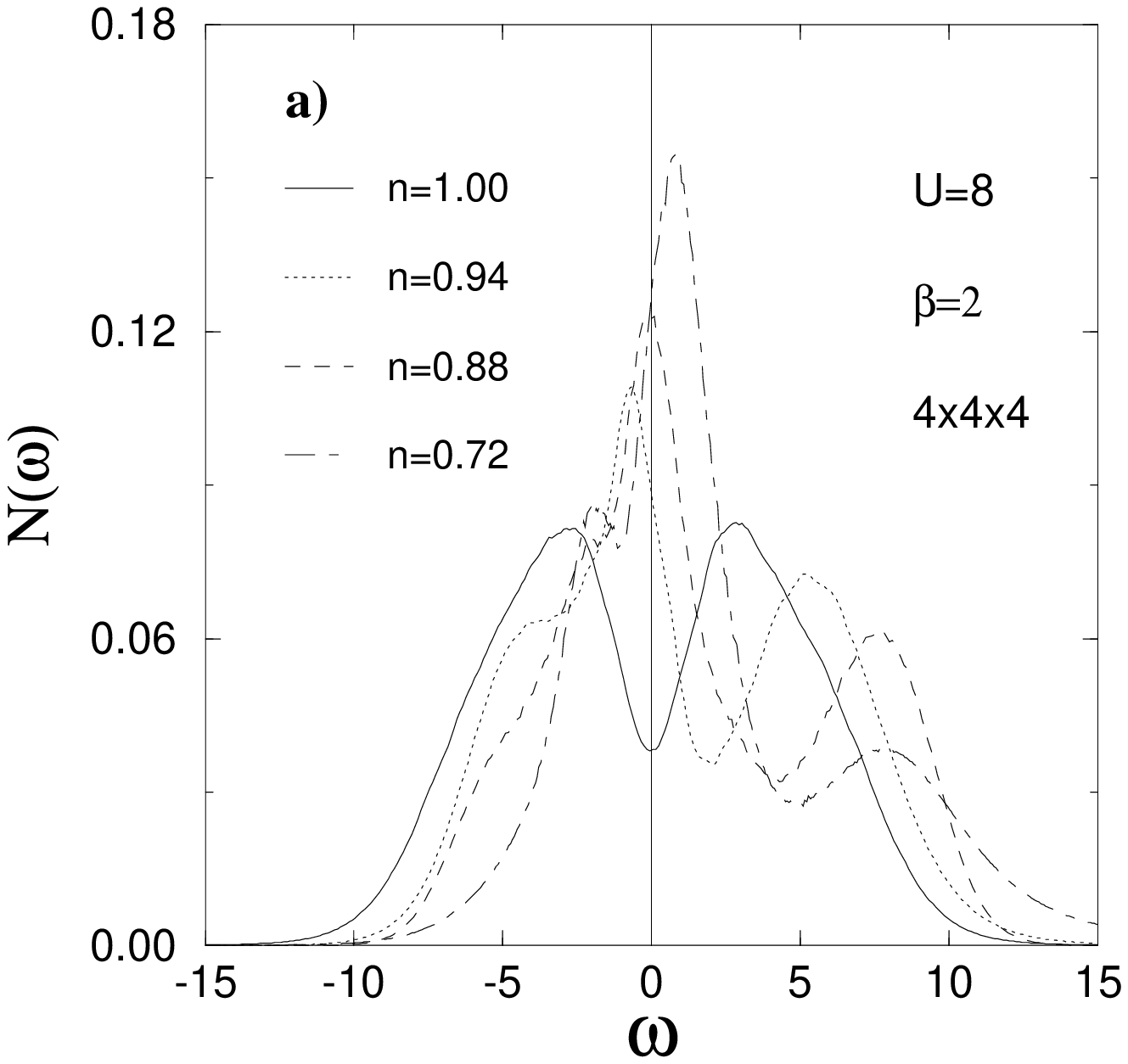,height=3.5in,width=4.2in}
\vskip-18mm
\psfig{file=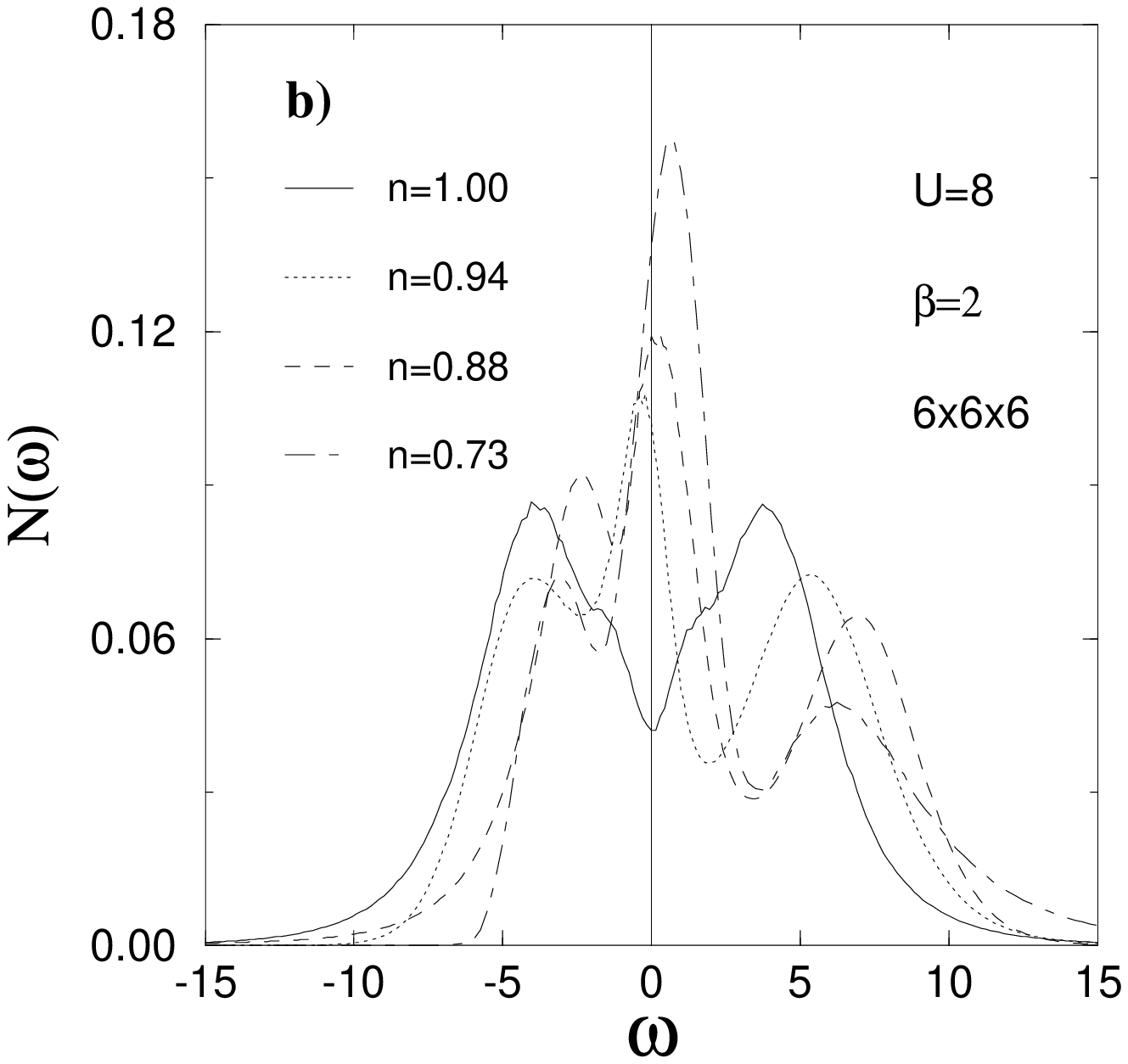,height=3.5in,width=4.2in}
\vskip-18mm
\caption{(a) $N(\omega)$ for the 3D Hubbard model on a $4^3$ lattice with
$U=8$ and  $\beta=2$ at several densities indicated in the figure;
(b) same as (a) but on a $6^3$ lattice. Frequencies are always relative 
to the chemical potential $\mu$.}

\end{figure}

Another important quantity to study is the quasiparticle residue Z.
The SCBA results 
show that Z is small but finite for the
case of one hole in an antiferromagnetic insulator state, and actually 
the results are
very similar in 3D and 2D systems.\cite{born,sasha} Numerical results
provide a similar picture.\cite{review}
On the other hand, Z vanishes in the
$D=\infty$ approach working in the paramagnetic state as the doping
$\delta$ tends to zero. Note that in this state there are no AF correlations
($\xi_{AF} =0$). Thus, it is clear that the hole quasiparticle at
half-filling observed in the 2D and 3D systems is $not$ related with the
quasiparticle-like feature observed in the PM state at $D=\infty$.
\begin{figure}
\psfig{file=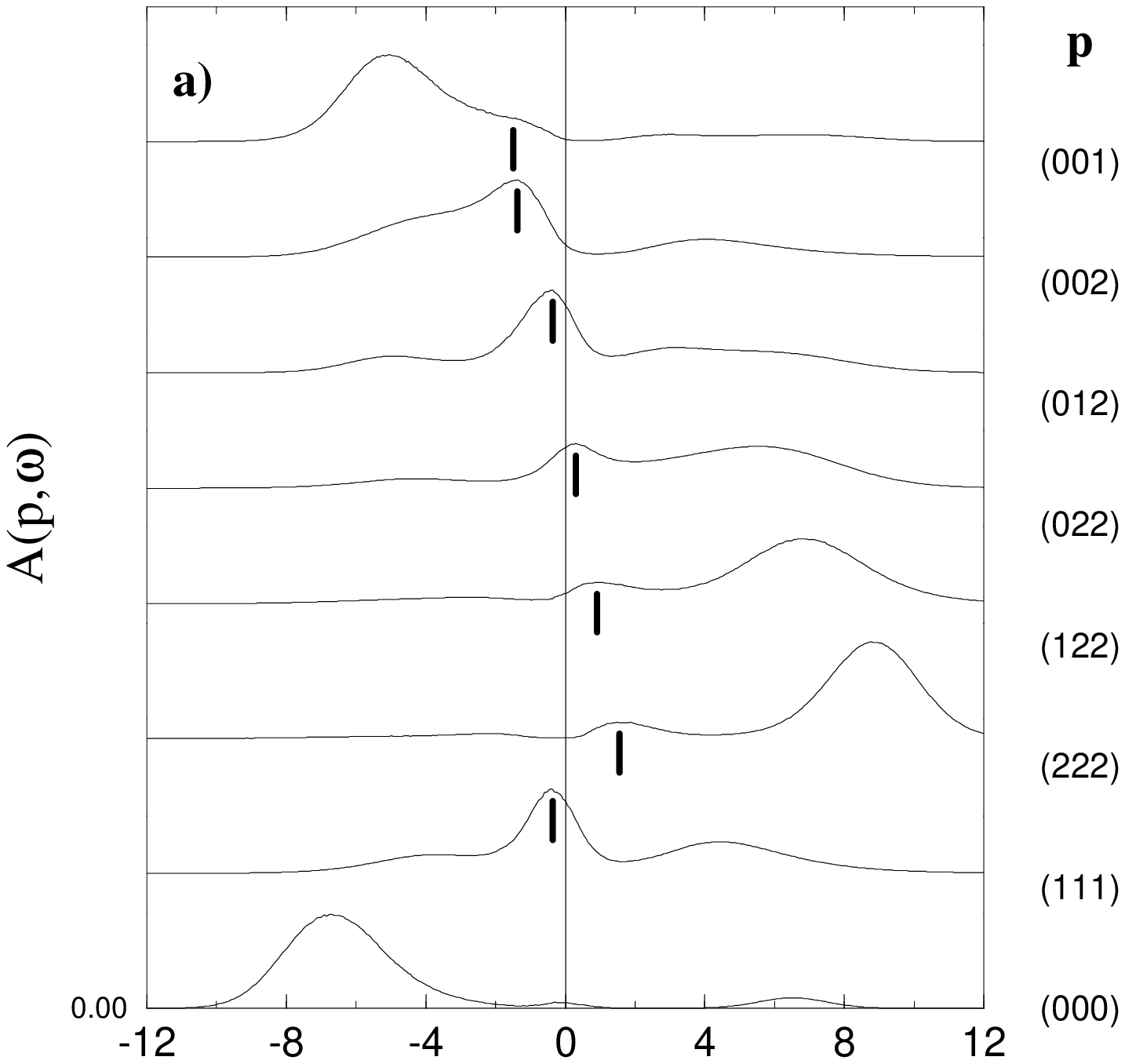,height=2.5in,width=4.0in}
\vskip-12mm
\psfig{file=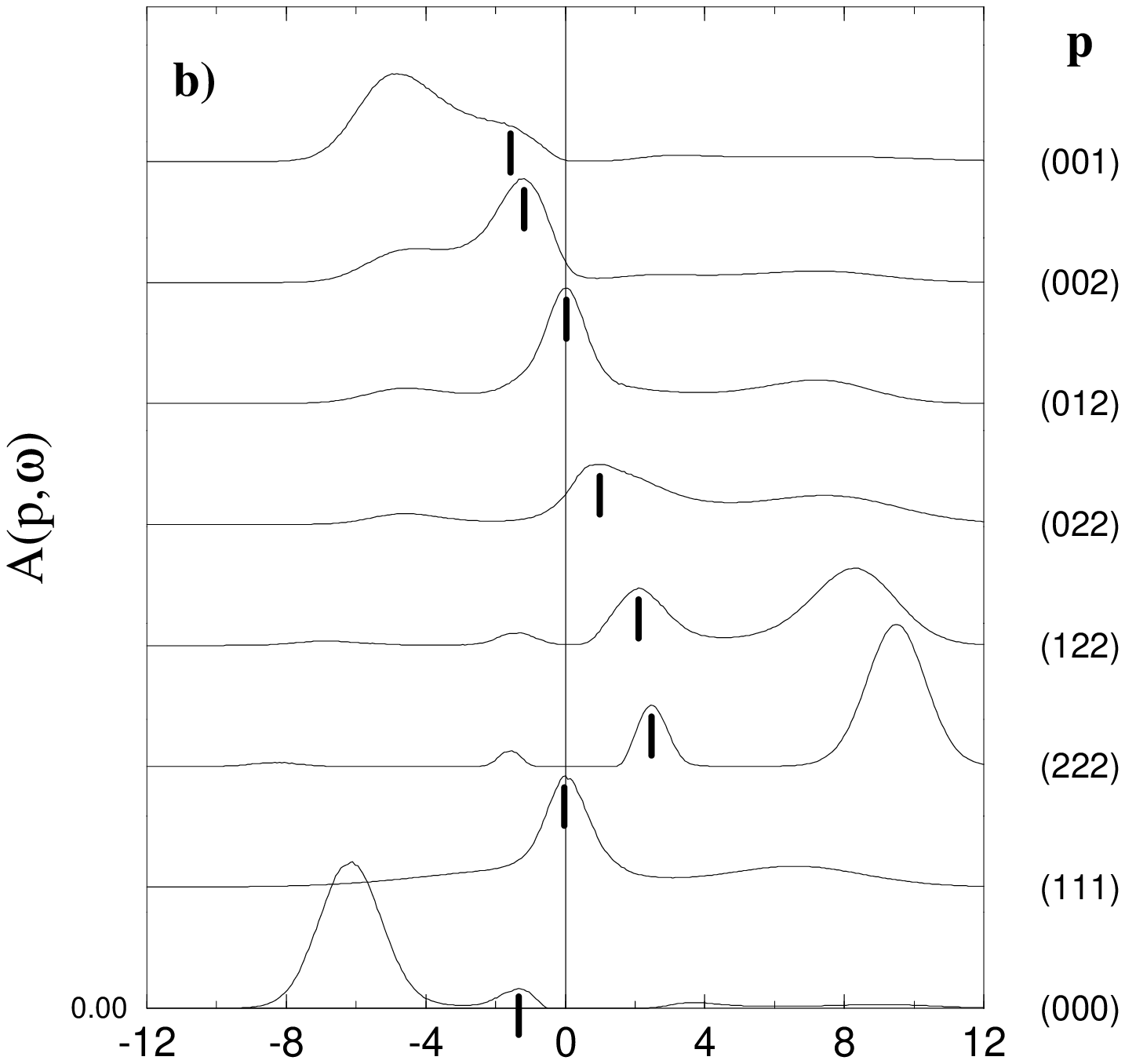,height=2.5in,width=4.0in}
\vskip-12mm
\psfig{file=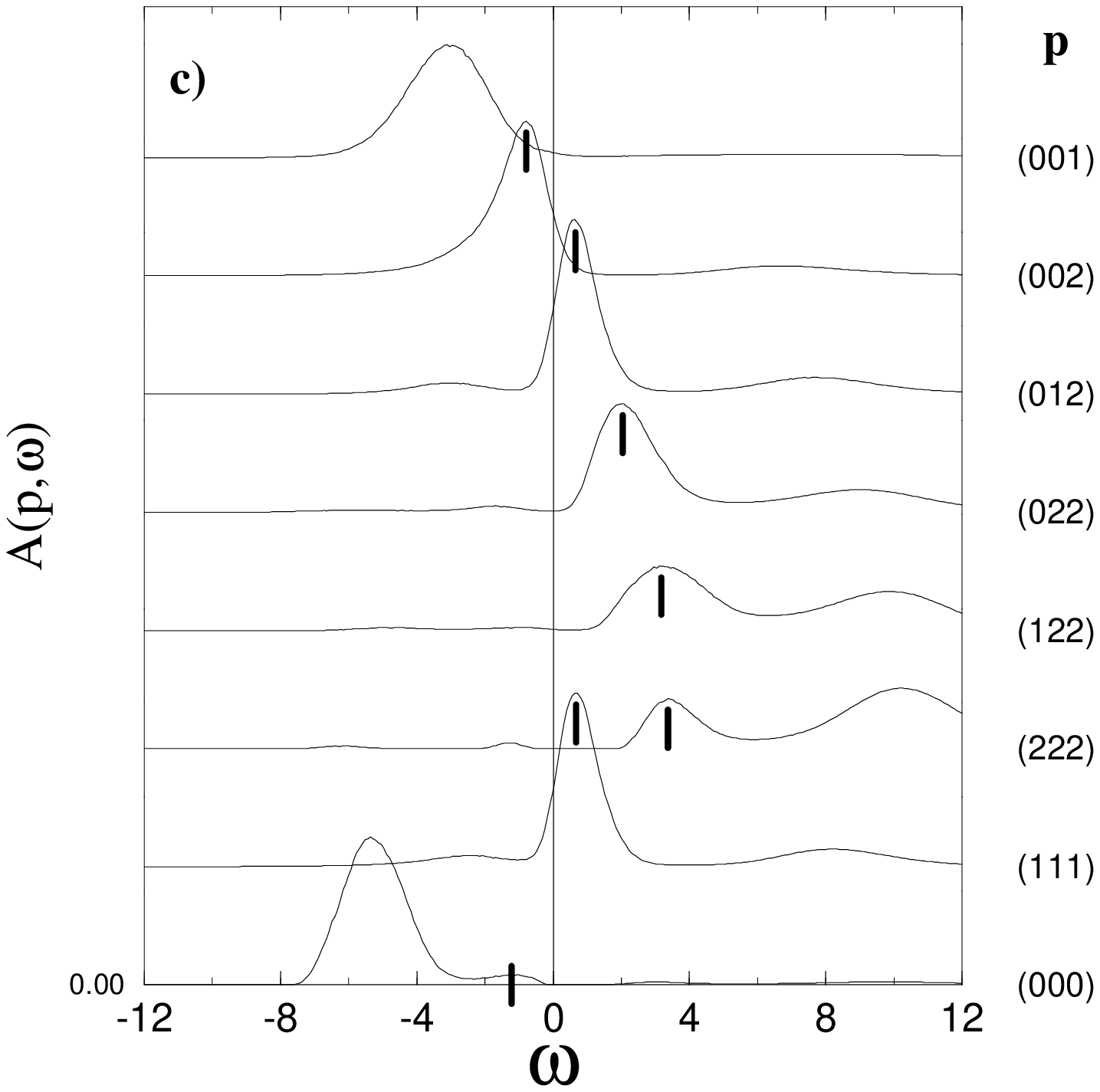,height=2.5in,width=4.0in}
\vskip-6mm
\caption{(a) Spectral weight $A({\bf p},\omega)$ of the 3D Hubbard model
calculated using QMC/ME 
on a $4^3$ lattice, at $U=8$,  $\beta=2$, and at density $\langle n \rangle
= 0.94$;
(b) same as (a) but at $\langle n \rangle = 0.88$;
(c) same as (a) but at $\langle n \rangle = 0.72$.}

\end{figure}

In Fig.~9, we show $A({\bf p},\omega)$ obtained on the $4^3$ lattice,
$U=8$, $T=1/2$ and various densities away from half-filling. 
The gap is now absent. From the energy location of the maximum of the
dominant peak in Fig.~9a-c, the quasiparticle dispersion can be obtained.
The results are shown in Fig.~10. It is remarkable that the quasiparticle
dispersion resembles that of a noninteracting system i.e. 
$\epsilon_{\bf p} = -2t^\star (\cos p_x + \cos p_y + \cos p_z)$, with a scale 
increasing from $t^\star \sim t/4$ to $t/3$ with doping. This dispersion 
certainly does not exhaust all the spectral weigth but a large 
incoherent part still remains 
at this coupling, density and temperature.
Similar results were observed in 2D.\cite{bulut,moreo,dos,ortolani} 
Only vestiges remain of the AF
induced weight in PES near $(\pi,\pi,\pi)$. However, this drastic
reduction of the AF induced intensity may be caused
by the high temperature of the simulation as observed in the
spin-spin correlation function (Fig.~7).

\begin{figure}

\vskip-5mm
\hspace*{7mm}
\psfig{file=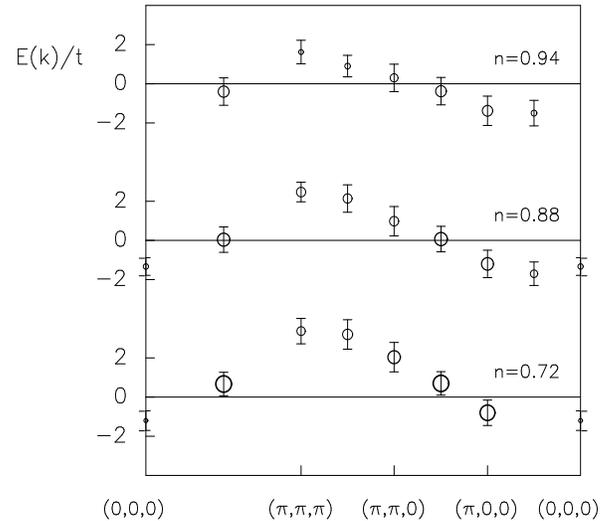,height=3.8in,width=3.5in,angle=-90}

\vskip-10mm
\caption{Dispersion of the dominant peak of Fig.~9a-c against momentum.
The densities  are indicated. The area in the circles is proportional to 
the peak intensity. Error-bars correspond to the half-width
of the peaks in the spectral weight.}

\end{figure}

\subsection{ $D=\infty$}

The previous subsection and the results at half-filling 
have shown that the DOS of the 3D Hubbard model has a large peak at the
top of the valence band.  The
peak is crossed by the chemical potential as $\langle n \rangle $
decreases.  This behavior is in apparent contradiction with results
reported at $D=\infty$ where a peak is generated upon doping if the
``paramagnetic'' solution to the mean-field problem is selected. At
$D=\infty$, there are only two very distinct magnetic ground states. One
has AF long-range order, and the other is a paramagnet with strictly
$zero$ AF correlation length i.e. without short range antiferromagnetic
fluctuations. Thus, at $D=\infty$ the transition is abrupt from a
regime with $\xi_{AF} = \infty$ to $\xi_{AF} = 0$. This does not occur
in finite dimensions where before the long-range order regime is
reached, AF correlations start building up smoothly. 
This qualitative difference is depicted in Fig.~11. 
\begin{figure}
\hspace*{5mm}
\psfig{file=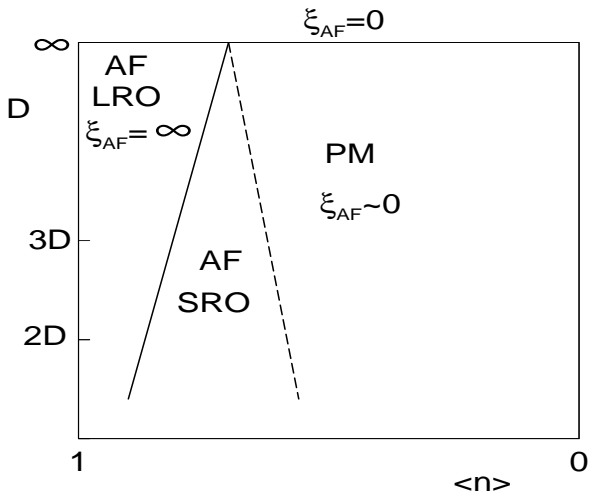,height=2.0in,width=2.5in,angle=-90}
\vskip20mm
\caption{``Phase diagram'' of the Hubbard model (1) in the 
$D$-$\langle n\rangle-$plane.
Solid line: AF phase boundary; dashed line: crossover where 
short range AF correlations disappear. The intermediate regime
of short range AF order vanishes in the limit $D\to\infty$.}

\end{figure}
$\xi_{AF}$ as small as a couple of lattice spacings can be robust
enough to induce important changes in the carrier dispersion, and may
even be enough to induce superconductivity as many theories for the
2D high-Tc cuprates conjecture. 
We believe that the absence of a regime of intermediate size AF
correlations at large D is the key ingredient that explains the 
differences reported here between D=2,3 and ${\rm D=\infty}$.

In Fig.~12a, the D=$\infty$
DOS in the AF phase is shown at $\langle n \rangle =1$
and 0.94. For these densities the AF-phase is energetically stable. We
observe the tendency of the large peak at the bottom of the valence band
to move towards the chemical potential in good agreement with the
3D Quantum Monte Carlo simulations. As found in 2D, the intensity of the
peak decreases as we move away from half-filling if the temperature is
low enough.
In Fig.~12b, the DOS in the $D=\infty$ limit working in the paramagnetic
phase is shown at several densities. 
For the present interaction, $U=8$, the paramagnetic solution remains 
metallic at all temperatures even at half filling. \cite{jarrell}
The results are qualitatively different from those observed in the AF regime.
At $\langle n \rangle =1$ a large peak at the chemical potential is clearly 
visible. Upon hole doping this peak gradually moves toward higher energies. 
At sufficiently strong doping the DOS of the PM-phase (Fig.~12b) resembles 
the results for the $3D$ lattices (Fig.~8), which is not surprising since 
AF correlations in $3D$ are strongly suppressed at the present temperature.
Close to half filling, however, the $3D$ results are closer to the DOS of 
the AF-phase where a strong peak is observed on the left hand side
of the chemical potential $\mu$. 
This result is gratifying since
the proper way to compare $D=3$ and $\infty$ results is by using
the actual ground states in each dimension.
In $D=\infty$, at low temperatures, 
the crossing of the peak by $\mu$ is expected at that point where the
AF-Phase becomes unstable against doping. 
\begin{figure}
\psfig{file=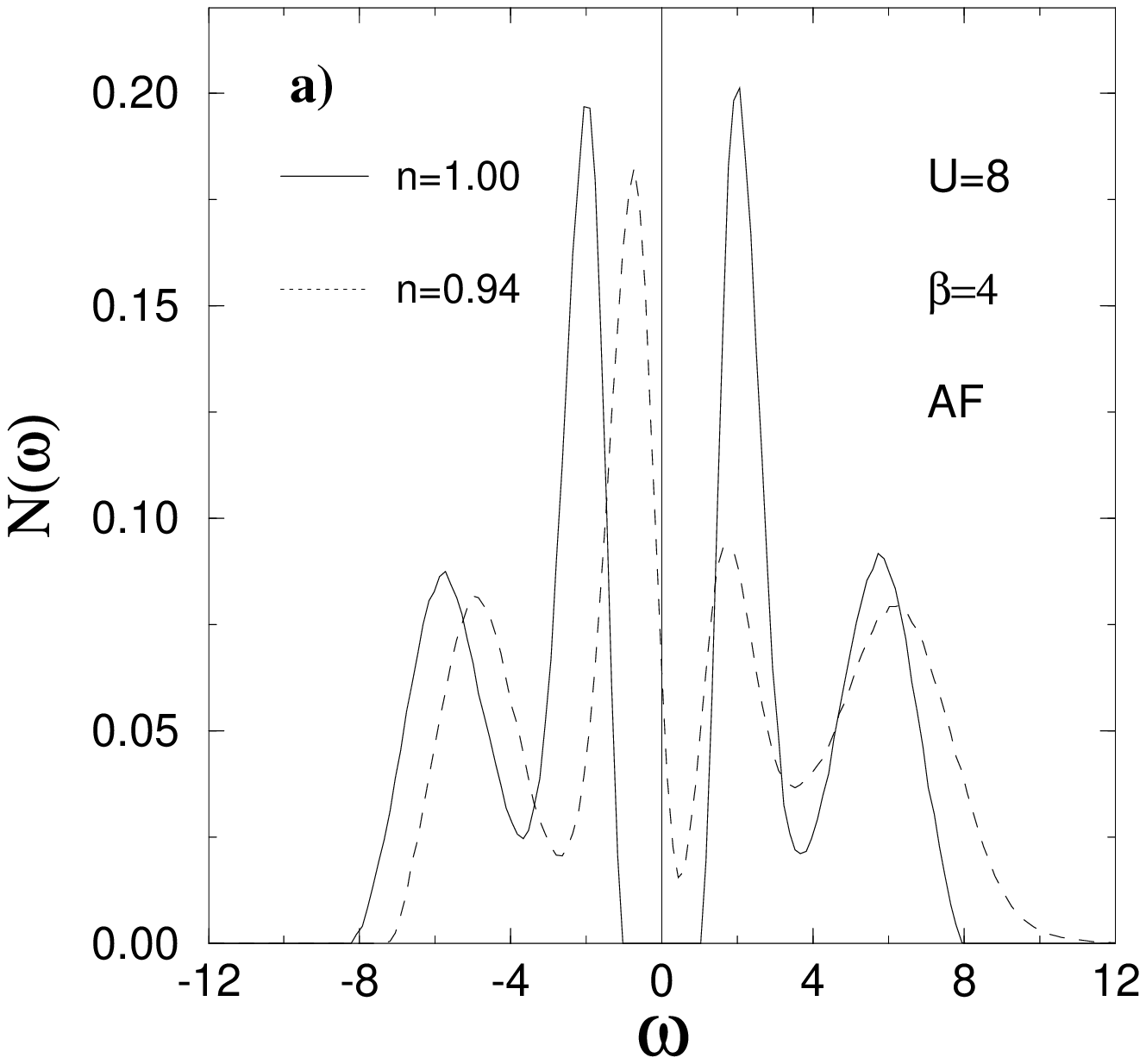,height=3.5in,width=4.2in}
\vskip-18mm
\psfig{file=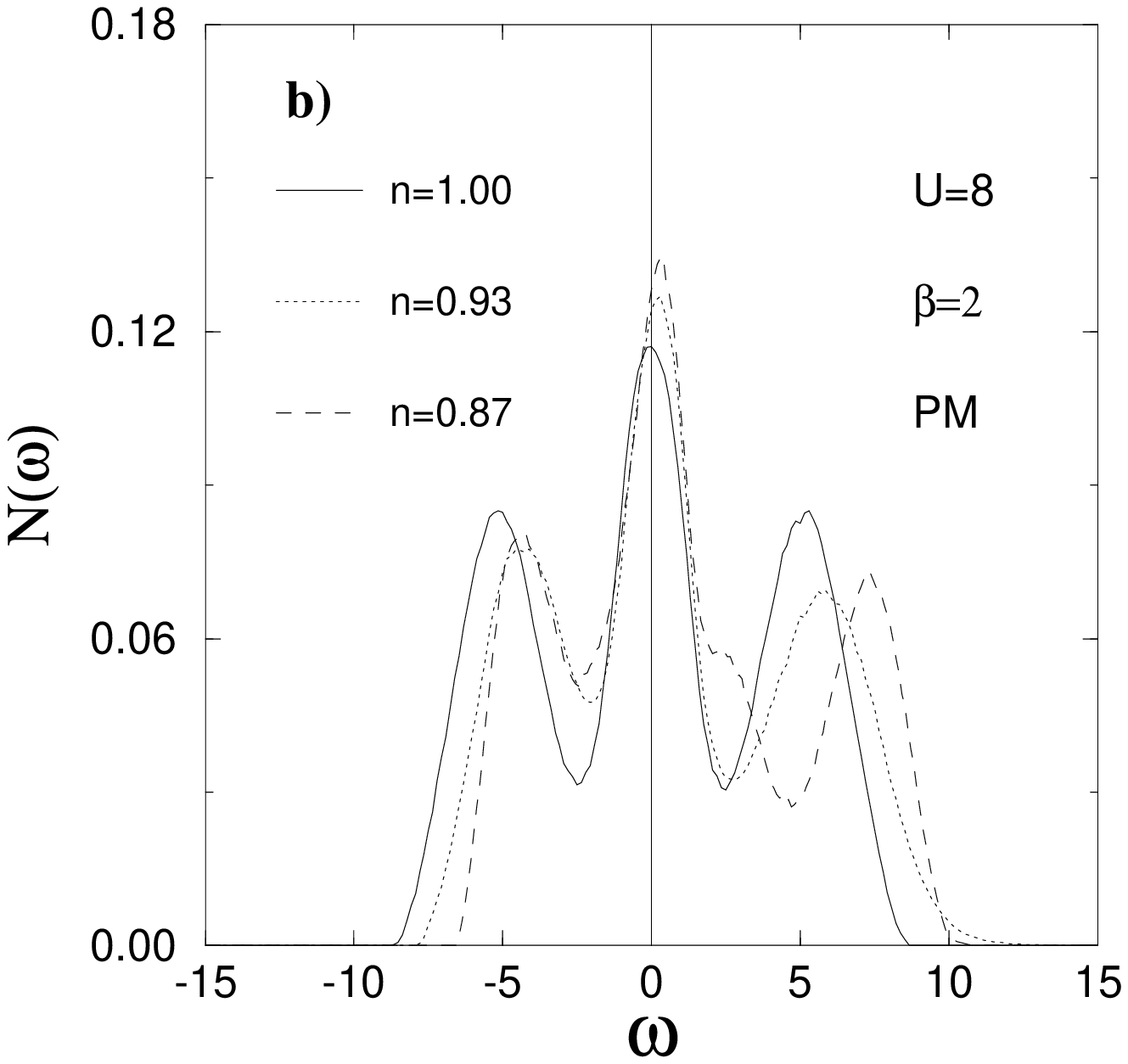,height=3.5in,width=4.2in}
\vskip-18mm
\caption{(a) $N(\omega)$ corresponding to the $D=\infty$ Hubbard model
at $U=8$,  $\beta=4$ at the electronic densities indicated.  The results
were obtained using the AF solution to the mean-field equation; (b)
same as (a) but using  $\beta=2$ and the PM solution to the mean-field
equation.}
\end{figure}

\section{\bf Conclusions}

In this paper we have calculated the single particle properties of the 3D
single band Hubbard model using Quantum Monte Carlo 
and the SDW mean field and SCBA approximations.
Our results have many similarities with those reported previously in
2D systems. 
At half-filling, peaks at the top of the valence band and bottom 
of the conduction band are observed in the DOS.
Their behavior is associated with spin polarons with a
bandwidth of order the exchange $J$. 
We found similarities to and semi-quantitive agreement with experimentally
observed features in the spectra of strongly correlated 3D AF insulators, 
$\rm{LaFeO_3}$ and $V_2O_3$.

As we dope the system, the sharp peak
associated with these quasiparticles is crossed by the chemical
potential as the density $\langle n \rangle $ changes. 
The PES weight observed away from half-filling is already present at
half-filling.  No new states are generated by doping.
This result must be contrasted with that observed experimentally in,
e.g., ${\rm Y_{1-x} Ca_x Ti O_3}$
using angle-integrated PES. In this case
spectral weight which is not present in the insulator
appears at $E_F$ in the metallic regime as we dope the system.
This behavior does not seem reproduced by the single band
Hubbard model Eq.(1) in 3D, whose physics appears to be very close to that of
2D. Indeed for the 2D cuprates it has been shown
experimentally that the states found at $E_F$ upon doping are already
present at half-filling.\cite{fujimori3}

An exception among 3D materials is $\rm{NiS_{2-x}Se_x}$ which remains 
antiferromagnetic throughout the metal-insulator transition induced by 
(homovalent) Se-substitution or temperature.
PES spectra at $x=0.5$ for different temperatures \cite{matsuura}
show a strong peak close to the Fermi energy which does $not$ disappear 
in the insulator. Instead the peak is shifted off the Fermi energy and 
only very slightly reduced in weight. Since this situation is not described
within the paramagnetic $D=\infty$ approach, AF correlations are presumably 
essential for the low energy electronic excitations of this system.

The success of the $D=\infty$
approach to the Hubbard model in describing the physics of
${\rm Y_{1-x} Ca_x Ti O_3}$, ${\rm SrVO_3}$ and ${\rm CaVO_3}$,
however, appears crucially to depend upon forcing the paramagnetic solution
of the equations.\cite{rozen1}  In this case,
states are actually $generated$ in the Hubbard gap
after a small hole doping is introduced.
Of course, it may be that the ``arbitrary'' choice of this 
paramagnetic solution, which is not the actual minimum of the 
free energy, is well motivated since it mimics the
presence of physical effects like frustration which
destroy long range order in real materials. More work is needed to show 
that this scenario is realized for realistic densities and couplings.

An alternative explanation for the discrepancy between
the PM solution in 
infinite-D and finite dimensional results may lie in the
finite resolution of the combination of Monte Carlo simulations 
and Maximum Entropy techniques.
However, the SCBA and results at
half-filling and low T show that it is likely that at $\langle n \rangle
=1$ we have quasiparticle states in the DOS. 

In studies of the single band Hubbard Hamiltonian
in 2D, and in the present analysis
in 3D, it is clear that short-range AF correlations play an important
role close to $\langle n \rangle =1$. In particular, the states created at the top
of the valence band are likely to be spin polarons with a finite
quasiparticle residue Z.  PES states observed 
at finite hole doping evolve continuously from
those present at half-filling.  Experiments on the 2D
high-Tc cuprates seem to present similar features, while the results for
the 3D perovskites are very different in the sense that
no remnants of the coherent part of the spectrum away from half-filling
are reported at half-filling.

Still, strong AF correlations are apparently present in several 3D transition
metal oxides and influence the low energy spectrum at least on the
insulating side of the transition. The introduction of frustration in the
single band Hubbard model in 3D, perhaps through next-nearest neighbor 
hoppings, will reduce AF correlations and in particular the AF-induced 
charge gap, and might be sufficient to observe a evolution of spectral
weight upon doping closer to the experimental findings. However, it might
be that 3D models which explicitly include orbital degeneracy 
will be necessary to reproduce the physics of the transition metal oxides,
as has recently been described for NiO chains\cite{nio} 
and Mn-oxides.\cite{muller}.
Indeed, a recent argument presented by Kajueter et al.\cite{kajueter} to 
justify the use of the $D=\infty$ model provides a more realistic 
explanation for the apparent link between theory in this limit,
and 3D transition-metal oxide results. The idea is that the
physics of the real perovskite 3D oxides is influenced by the orbital
degeneracy. Presumably this effect leads to a drastic reduction of the
antiferromagnetic correlations that dominate the
physics of these 2D and 3D systems. 
Many orbitals, including Hund's coupling, produce an effective magnetic 
frustration that may reduce the AF correlation length to a 
negligible value even close to the 
AF insulator at half-filling. Such a frustration effect could be strong enough 
to generate a finite critical coupling $U/t$ at half-filling. 

\section{Acknowledgments}
\medskip 
We thank A.~Fujimori, M.~Rozenberg and A.~Moreo for useful discussions. 
We are grateful to A.~Sandvik for providing his Maximum Entropy program.
Most simulations were done on a cluster of HP-715 work stations at the
Electrical and Computer Engineering Department at UC Davis.
We thank P.~Hirose and K.~Runge for technical assistance.
E.~D.~is supported by grant NSF-DMR-9520776.
R.~T.~S.~is supported by grant NSF-DMR-9528535.
M.~U.~is supported by a grant from the Office of Naval Research,
ONR N00014-93-1-0495 and by the Deutsche Forschungsgemeinschaft.
We thank the National High Magnetic Field Laboratory (NHMFL) and the
Center for Materials Research and Technology (MARTECH) for additional support.
\medskip


%


\begin{references}

\bibitem[*]{} New address:
Theoretische Physik III, Universit\"at Augsburg, D--86135 Augsburg, Germany.

Electronic address: ulmke@physik.uni-augsburg.de

\bibitem{HUBBARD} M.C. Gutzwiller, Phys. Rev. Lett. {\bf 10}, 159 (1963);
J. Hubbard, Proc. Roy. Soc. {\bf A 276}, 238 (1964);
J. Kanamori, Prog. Theor. Phys. {\bf 30}, 275 (1963).

\bibitem{review} E. Dagotto, Rev. Mod. Phys. {\bf 66}, 763 (1994),
and references cited therein.

\bibitem{flat-exper} 
D.S. Dessau et al., Phys. Rev. Lett. {\bf 71}, 2781 (1993);
K. Gofron et al., Phys. Rev. Lett. {\bf 73}, 3302 (1994).

\bibitem{flat} E. Dagotto, A. Nazarenko and M. Boninsegni,
Phys. Rev. Lett. {\bf 73}, 728 (1994).

\bibitem{bulut} N. Bulut, D.J. Scalapino and S.R. White, Phys. Rev.
{\bf B50}, 7215 (1994).

\bibitem{hanke} R. Preuss, W. Hanke, and W. von der Linden, Phys. Rev. 
Lett.  {\bf 75}, 1344 (1995).

\bibitem{berlin} M. Langer, J. Schmalian, S. Grabowski, and K.H. 
Bennemann, Phys. Rev. Lett. {\bf 75}, 4508 (1995).

\bibitem{fujimori} A. Fujimori, I. Hase, H. Namatame, Y. Fujishima, Y.
Tokura, H. Eisaki, S. Uchida, K. Takegahara, and F. M. F. de Groot,
Phys. Rev. Lett. {\bf 69}, 1796 (1992).

\bibitem{inoue} I. H. Inoue, I. Hase, Y. Aiura, A. Fujimori, Y.
Haruyama, T. Maruyama, and Y. Nishihara, Phys. Rev. Lett. {\bf 74}, 2539
(1995).

\bibitem{morikawa} K. Morikawa, T. Mizokawa, K. Kobayashi, A. Fujimori,
H. Eisaki, S. Uchida, F. Iga, and Y. Nishihara, Phys. Rev. {\bf B 52},
13711 (1995).

\bibitem{WHITE}
M. Vekic and S.R. White, Phys. Rev. {\bf B47}, 1160 (1993);
G.S. Feng and S.R. White, Phys. Rev. {\bf B46}, 8691 (1992); and
M. Vekic and S.R. White, Phys. Rev. {\bf B47}, 5678 (1992).

\bibitem{jarrell} M. Jarrell and T. Pruschke, Z. Phys. {\bf B 90}, 187
(1993).

\bibitem{fujimori2} A. Fujimori, I. Hase, M. Nakamura, H. Namatame, Y.
Fujishima, Y. Tokura, M. Abbate, F. M. F. de Groot, J. C. Fuggle, O.
Strebel, M. Doce, and G. Kaindl, Phys. Rev. {\bf B 46}, 9841 (1992).

\bibitem{tokura} Y. Tokura, Y. Taguchi, Y. Okada, Y. Fujishima, T.
Arima, K. Kumagai, and Y. Iye, Phys. Rev. Lett. {\bf 70}, 2126 (1993).

\bibitem{metzner} W. Metzner and D. Vollhardt, Phys. Rev. Lett. {\bf 62}, 324 
(1989); for a review see D. Vollhardt, in {\it Correlated Electron Systems},
ed. V. J. Emery (World Scientific, Singapore, 1993), p. 57.

\bibitem{rrmp}  For a review see 
A. Georges, G. Kotliar, W. Krauth, and 
M. Rozenberg, Rev. Mod. Phys. {\bf 68} 13 (1996).

\bibitem{georges}  For a review see Th. Pruschke, M. Jarrell, 
and J. K. Freericks, Adv. Phys. {\bf 44}, 187 (1995).

\bibitem{blankenbecler} R. Blankenbecler, D. J. Scalapino, R. L. Sugar,
 Phys. Rev. {\bf D 24}, 2278 (1981).

\bibitem{white} S. R. White, D. J. Scalapino, R. L. Sugar, E. Y. Loh,
J. E. Gubernatis, R. T. Scalettar, Phys. Rev. {\bf B 40}, 506 (1989).

\bibitem{hirsch} J. E. Hirsch, Phys. Rev. {\bf B 28}, 4059 (1983).

\bibitem{HIRSCHn} J. E. Hirsch, Phys. Rev. {\bf B 35}, 1851 (1987).

\bibitem{rts} R. T. Scalettar, D. J. Scalapino, R. L. Sugar and D.
Toussaint, Phys. Rev. {\bf B 39}, 4711 (1989).

\bibitem{gubernatis}  For a review see M. Jarrell and J. E. Gubernatis, 
Phys. Rep. {\bf 269}, 133 (1996).

\bibitem{bethe} We choose a half-elliptic, non-interacting density of states
which becomes exact for the Bethe lattice with infinite connectivity.
Like real $3D$-DOS and in contrast to the hypercubic lattice in $D=\infty$ 
it has a finite bandwidth $(W=4t^*)$ and algebraic band edges.

\bibitem{fye} J. E. Hirsch and R. M. Fye, 
Phys. Rev. Lett. {\bf 56}, 2521 (1986).

\bibitem{rspec} M. Rozenberg, G. Kotliar, and H. Kajueter, preprint.

\bibitem{moreo} A. Moreo et al.,
Phys. Rev. {\bf B 51}, 12045 (1995).

\bibitem{schrieffer} A. Kampf and J. R. Schrieffer, Phys. Rev. {\bf B
41}, 6399 (1990).

\bibitem{bulut1} N. Bulut, D. J. Scalapino, and S. R. White, Phys. Rev.
Lett. {\bf 73}, 748 (1994).

\bibitem{shadow} P. Aebi et al., Phys. Rev. Lett. {\bf 72}, 2757 (1994);
S. Haas et al., Phys. Rev. Lett. {\bf 74}, 4281 (1995).

\bibitem{wells} B.O. Wells, et al., Phys. Rev. Lett. {\bf 74}, 964 (1995).

\bibitem{sarma} D. D. Sarma et al., Phys. Rev. {\bf B 49}, 14238 (1994).

\bibitem{shin} S. Shin et al., J. Phys. Soc. Jpn. {\bf 64}, 1230 (1994).

\bibitem{rozen2}
M.J. Rozenberg, G. Kotliar, H. Kajueter, G.A. Thomas, D.H. Rapkine,
J.M. Honig, and P. Metcalf,
Phys. Rev. Lett. {\bf 75} 105, 1995.

\bibitem{sdw} J. R. Schrieffer, X. G. Wen, and S. C. Zhang, Phys. Rev.
{\bf B 39}, 11663 (1989).

\bibitem{renormu}
N. Bulut, D.J. Scalapino, and S.R. White, 
Phys. Rev. {\bf B47}, 14599 (1993).

\bibitem{vandongen}
P. van Dongen, Phys. Rev. Lett. {\bf 67}, 757 (1991). 
The self-consistent perturbation theory around the SDW MF solution
provides a renormalization factor about $q=0.29$ in the limit of small
$U$ in $D=3$ and $D=\infty$. Numerically it was found that $q$ increases
with $U$, see e.g.~\cite{jarrell}. 

\bibitem{born} 
S. Schmitt-Rink, C.M. Varma, and A.E. Ruckenstein,
Phys. Rev. Lett. {\bf 60}, 2793 (1988); 
F. Marsiglio, A.E. Ruckenstein,S. Schmitt-Rink, and C.M. Varma, Phys. Rev. 
{\bf B43}, 10882 (1991);
G. Martinez and P. Horsch, Phys. Rev. {\bf B44}, 317 (1991);
Z. Liu and E. Manousakis, Phys. Rev. {\bf B45}, 2425 (1992).

\bibitem{sasha} A. Nazarenko and E. Dagotto, preprint.

\bibitem{bulut2} N. Bulut, D.J. Scalapino and S.R. White, Phys. Rev. 
Lett. {\bf 72}, 705 (1994).

\bibitem{dos} A. Nazarenko, S. Haas, J. Riera, A. Moreo, and E. Dagotto, 
preprint.

\bibitem{dos1}
Recently proposed theories of high-Tc make extensive
use of a large accumulation of weight in the DOS near the chemical potential
to enhance the critical temperature 
[E. Dagotto, A. Nazarenko and A. Moreo,
Phys. Rev. Lett. {\bf 74}, 310 (1995)].

\bibitem{ortolani} E. Dagotto, F. Ortolani and D. Scalapino,
Phys. Rev. {\bf B 46}, 3183 (1992).

\bibitem{fujimori3} A. Fujimori et al., Phys. Rev. {\bf B 39}, 2255
(1989); Phys. Rev. {\bf B 40}, 7303 (1990).

\bibitem{matsuura} A. Y. Matsuura, Z.-X. Shen, D. S. Dessau, C.-H. Park,
T. Thio, J. W. Bennett, O. Jepson,  Phys. Rev. {\bf B 53}, R7584 (1996).

\bibitem{rozen1}
M. J. Rozenberg , I. H. Inoue , H. Makino , F. Iga , Y.

\bibitem{nio} E. Dagotto, J. Riera, A. Sandvik, and A. Moreo, Phys. Rev.
Lett. {\bf 76}, 1731 (1996).

\bibitem{muller} E. M\"uller-Hartmann and E. Dagotto, preprint.

\bibitem{kajueter} H. Kajueter, G. Kotliar, and G. Moeller, preprint
Rutgers Univ. (1996).

\end{references}
\end{document}